\documentclass[aps,pra,twocolumn,showpacs,superscriptaddress,groupedaddress]{revtex4}
\usepackage[latin1]{inputenc}
\usepackage{bm}
\usepackage{multirow,amssymb,amsbsy,amsmath}
\usepackage{graphicx}
\usepackage{verbatim}
\makeatletter
\usepackage{pifont}
\makeatother

\begin{document}

\title{Efficient Quantum State Estimation with Over-complete Tomography}

\author{Chi Zhang}
\affiliation{Key Laboratory of Quantum Information, University of
Science and Technology of China, CAS, Hefei, 230026, P.R.China}

\author{Guo-Yong Xiang}
\affiliation{Key Laboratory of Quantum Information, University of
Science and Technology of China, CAS, Hefei, 230026, P.R.China}

\author{Yong-Sheng Zhang$\footnote{email:yshzhang@ustc.edu.cn}$}
\affiliation{Key Laboratory of Quantum Information, University of
Science and Technology of China, CAS, Hefei, 230026, P.R.China}

\author{Chuan-Feng Li$\footnote{email:cfli@ustc.edu.cn}$}
\affiliation{Key Laboratory of Quantum Information, University of
Science and Technology of China, CAS, Hefei, 230026, P.R.China}

\author{Guang-Can Guo}
\affiliation{Key Laboratory of Quantum Information, University of
Science and Technology of China, CAS, Hefei, 230026, P.R.China}

\date{\today}

\begin{abstract}
It is widely accepted that the selection of measurement bases can
affect the efficiency of quantum state estimation methods, precision
of estimating an unknown state can be improved significantly by
simply introduce a set of symmetrical measurement bases. Here we
compare the efficiencies of estimations with different numbers of
measurement bases by numerical simulation and experiment in optical
system. The advantages of using a complete set of symmetrical
measurement bases are illustrated more clearly.
\end{abstract}

 \maketitle

\section*{I. Introduction}

Quantum states are the most important and fundamental elements in
the quantum world. Because of the superposition property, quantum
states are highly complex and difficult to identify. Quantum state
tomography (QST) \cite{b1, b2}, a universal method to reconstruct
quantum states by making a series of measurements on an ensemble,
plays an important role in quantum information science \cite{a}.
However, because of the asymmetrical distribution of the measurement
bases of the early QST methods \cite{b1}, all random errors
(statistical or technological) \cite{d} accumulate during the
measurements so that the quantum state estimation is always
inefficient \cite{Wootters}. Then a new symmetrical tomography
scheme, called over-complete QST \cite{Kaltenbaek, Langford}, had
been established. Its efficiency has been proven and it has been
widely used in many experiments \cite{Adamson, Olmschenk, Lima, y1,
y2, y3, y4}. This method may find versatile usage in many fields,
such as quantum computation and simulation \cite{c1, c2}, quantum
communication \cite {a, c3}, quantum metrology \cite{c4, c5},
condensed matter physics \cite{c6}, quantum chemistry \cite{c7, c8},
and quantum biology \cite{c9, c10}.

In this letter, first, we analyze and numerically simulate the
quantum state estimation procedure and compare the efficiencies of
over-complete QST with that of normal QST, then we experimentally
demonstrate its efficiency in optical system.

\begin{figure}[tb]
\centering
\includegraphics[width=0.4\textwidth]{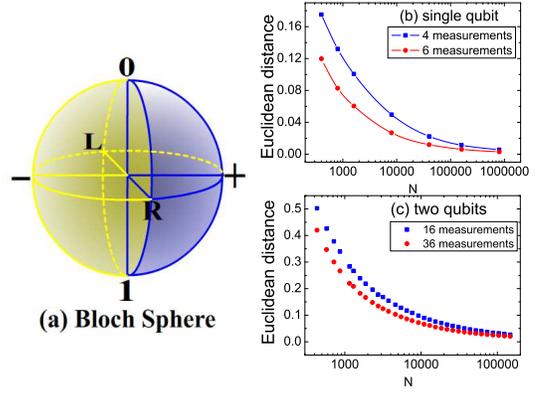}
\caption{\label{Fig:1}(color online) Errors of over-complete QST and
normal QST. (a). The measurement bases of over-complete QST (yellow
and blue) are distributed symmetrically on the surface of the Bloch
Sphere, while those of normal QST (blue) only cover a quarter of the
surface. In (b) and (c), 1000 different states are chosen randomly
from the entire state space. For each state, we simulate the
tomography procedure with $4^n$ (blue) and $6^n$ (red) measurement
bases for one qubit (b) and two qubits (c), where $n$ is the number
of qubits, with the same total resources $N$. We vary $N$
approximately from $10^2$ to $10^6$, and calculate the error of
estimation by Euclidean distance for each chosen value of $N$. Each
point represents the mean error of estimation for the 1000 states at
each different $N$. One thousand samples are sufficient to draw the
curve because the mean value and error bar do not change much
between data volumes of 500 and 1000. It is obvious that
over-complete QST is more precise. }
\end{figure}

\section*{II. Numerical Simulation}

Suppose that there are $N$ copies of a quantum state, which is
totally unknown before the measurements are made for the
reconstruction task. An $n$-qubit quantum state can be described by
a $2^n$ dimensional density matrix \cite{g}. Because the density
matrix has the properties of normalization, Hermiticity, and
positivity, it contains $4^n-1$ variables, so we need at least $4^n$
measurements to determine the density matrix of the state. The
standard QST method to reconstruct the density matrix is to equally
divide the $N$ copies into exactly $4^n$ measurements, record the
counts $n_\nu$ for every measurement, and then apply maximum
likelihood estimation (MLE) \cite{b1} to find the physical quantum
state that is most likely to generate these measured data. To be
specific, these measured data are assumed to be Gaussian distributed
around the true values, so the probability of observing a particular
set of counts is given by
\begin{equation} \label{eq1}
P(n_1,...n_m)=\frac{1}{N_{norm}}exp[-\sum_{\nu=0}^{m} \frac{(\bar
n_\nu-n_\nu )^2}{2 \bar n_\nu}],
\end{equation}
where $N_{norm}$ is the normalization constant, $m$ is the number of
measurements, and $\bar n_\nu=\frac{N}{m}\langle \psi_\nu | \hat\rho
| \psi_\nu\rangle$ is the expectation value of $| \psi_\nu\rangle$
measuring the state $\hat\rho$. Therefore, the physical density
matrix with maximum $P(n_1,...n_m)$ is most likely to be the true
state. This is a universally applicable method to estimate any
previous unknown quantum states.

A serious problem of QST is that the selection of measurement bases
is always asymmetric in the state space; in the Bloch Sphere
\cite{h} of a single qubit, for example, the four bases in early
method cover only a quarter of the surface (Figure 1(a)). However,
using the least possible number of measurement bases does not
necessarily cost the minimum total resources. The real difficulties
in experiments, such as the estimation of a quantum state of eight
ions \cite{f}, are the limited resources, not the measurement times.
According to previous study on the choice of measurement sets
\cite{pra}, inscribed Platonic solid of the Bloch sphere measurement
sets result in the minimum error of estimation. This indicates that
symmetrically arranged measurements will improve the precision
significantly, in this letter, we provide insightful research on why
the selection of measurement sets affects the precision and
experimentally demonstrate this point. By numerical simulation of
the tomography procedure on single-qubit and two-qubit states, it
becomes clear that when these $N$ copies are distributed equally
into $6^n$ symmetrical measurement bases, i.e., each measurement
base receives $N/6^n$ copies, the precision of the state estimation
will be improved significantly. The chosen bases are of the form
\begin{equation} \label{eq2}
|\psi\rangle=\otimes_{\nu=1}^n| \psi_\nu\rangle,
\end{equation}
where $| \psi_\nu\rangle\in \{|0\rangle$, $|1\rangle$, $|+\rangle$,
$|-\rangle$, $|L\rangle$, $|R\rangle$\}, with $|\pm\rangle =
\frac{1}{\sqrt 2}(|0\rangle \pm |1\rangle)$, $|L\rangle =
\frac{1}{\sqrt 2}(|0\rangle + i|1\rangle)$, $|R\rangle =
\frac{1}{\sqrt 2}(|0\rangle - i|1\rangle)$, which lie at the six
polar symmetrical points of the Bloch Sphere (FIG. 1(a)). MLE is
used to find the most likely density matrix, as in QST. The results
are shown in FIG. 1 (b) and (c), wherein the error of estimation is
represented by Euclidean distance $\sqrt {Tr[(\rho - \hat\rho)^2]}$
\cite{a}. As can be seen from the figure, over-complete QST needs
fewer resources than normal QST to reach the same precision.

In our work, the task of finding the minimum of likelihood function
is efficiently executed by the simulated annealing algorithm
\cite{z}, which proceeds as follows. (i). Start by estimating the
target matrix, $\hat\rho$, as an arbitrary physical density matrix,
such as the $I/4$, with an initial temperature $T_0$. (ii). Let the
current temperature $T$ decrease as the procedure continues, then
transform the matrix to another one (for instance, change an element
or partially transpose the matrix), and calculate the value of the
likelihood function. The probability of accepting the new $\hat\rho$
depends on the ratio of the new and the old likelihood functions,
and on the current temperature $T$; more precisely, it is
\begin{equation}
min\{exp[-\frac{L(\hat\rho_{n+1}) - L(\hat\rho_n)}{kT_n}], 1\},
\end{equation}
where $k$ is a constant number. (iii). Repeat step (ii) until the
value of the likelihood function becomes stable. (iv). Report
$\hat\rho$ as the estimated density matrix. Analogous to the
annealing process in solid state matter, the transition probability
between two levels is determined by the temperature and the energy
difference of the two levels. With $T$ decreasing, the system will
finally arrive at the ground state. In a similar way, the simulated
annealing algorithm can find the minimum quickly and precisely.

First, we analyze the situation of single qubit. Finding the density
matrix most likely to match the experimental data is very like the
procedure of fitting a line with the least squares method \cite{i}.
Obviously, at least two points are needed to determine a line. To
determine a line more accurately, we may measure each of the two
points with more resources to make the measurements more reliable,
or we may use a series of measurements and fit the line with more
points. The latter approach is better, and it is therefore always
used. The reason is that even though any single point is not so
accurate, their statistical errors, which are random, will cancel
each other. Moreover, if some types of random errors are related to
the selection of the measurement bases, they will accumulate in the
first method but cancel each other in the second. Especially when
the measurement resources are distributed symmetrically, these
errors will be expected to reduce to the minimum. In a similar way,
when estimating an unknown quantum state, the statistical error is
closely related to the count $n_\nu$, which is determined by the
measurement base. Technological errors, such as the uncertainty on
the angle of the wave plate, will also accumulate if there are only
$4$ measurement bases. Single qubit states can be described by
points on the Bloch Sphere, and measurement bases can only lie on
the surface. Therefore, the precision of estimation will be improved
significantly if we use the six polar points, which are distributed
symmetrically on the sphere, as measurement bases. In FIG. 1(a),
these measurement bases can be regarded as 6 directions from which
to observe the unknown states. Any particular set of 4 of them
cannot reach a fine estimation because they cannot see the state
from all directions (they cover only a quarter of the sphere), even
though they are enough to determine the state in theory. For more
than one qubit, the space of states grows exponentially, and we need
to use the combination of these bases described by Eq. (\ref{eq2})
to make up a complete set of symmetrical measurements.

\begin{figure}[tb]
\centering
\includegraphics[width=0.4\textwidth]{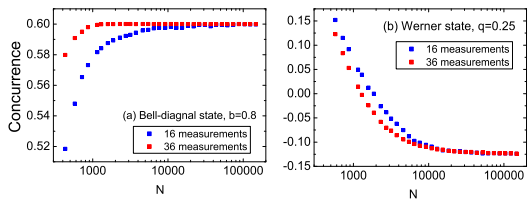}
\caption{\label{Fig:2}(color online) Error of concurrence
estimation. (a). Bell-diagonal state with b=0.8, the true
concurrence is 0.6. (b). Werner state with q=0.25, the true
concurrence is -0.125 (here we do not set its minimum value to
zero). $N$ is the total quantity of resources used for estimation.
These two examples illustrate that over-complete QST is always
closer to the true value than QST. We generate the curve with 300
samples for each point.}
\end{figure}

In addition, to show more advantages of the over-complete QST
method, we numerically simulate the estimation of concurrence
\cite{j}, an important parameter in two-qubit entanglement
verification \cite{j2}. It is defined as
\begin{equation} \label{eq3}
C(\rho)=max \{ 0, \lambda _1 -\lambda _2 -\lambda _3 -\lambda _4 \},
\end{equation}
where $\lambda _i$ are the eigenvalues, in decreasing order, of the
Hermitian matrix $R=\sqrt {\sqrt{\rho} \tilde{\rho} \sqrt{\rho}}$,
and $\rho$ is the density matrix. Here we do not set its minimum
value to zero so that we may study the estimation more clearly. Two
common states, the Bell-diagonal state \cite{k1}
\begin{equation} \label{eq4}
\rho = b | \psi^-\rangle \langle \psi^-| + (1-b) | \phi^-\rangle
\langle \phi^-|
\end{equation}
where $| \psi^-\rangle = \frac{1}{\sqrt 2} (|00\rangle
-|11\rangle)$, $| \phi^-\rangle = \frac{1}{\sqrt 2} (|01\rangle
-|10\rangle)$, and the Werner state \cite{k2}
\begin{equation} \label{eq5}
\rho=q|\psi^-\rangle\langle\psi^-|+(1-q)I/4
\end{equation}
are used as examples. The results are shown in FIG. 2. It is easy to
see from the figure that the concurrence estimated by 36
measurements is closer to the true value than that estimated by 16
measurements. When data are insufficient, the deviation seems
tremendous, the effect of which is caused by MLE \cite{e}.
Nonetheless, over-complete QST has reduced the deviation greatly.
Consequently, the result from over-complete QST is more reliable
than that obtained from normal QST, and over-complete QST is very
useful to verify entanglement when resources are limited.

\begin{figure}[tb]
\centering
\includegraphics[width=0.5\textwidth]{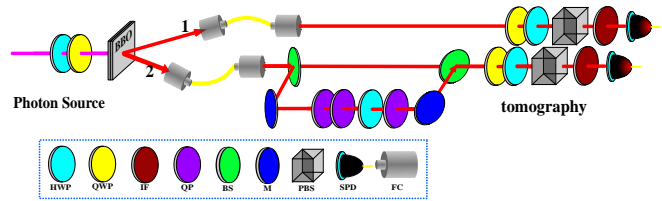}
\caption{\label{Fig:3} (color online) Experimental setup. Here, the
abbreviations of the components stand for the following: HWP -- half
wave plate, QWP -- quarter wave plate, IF -- interference filter, QP
-- long quartz plate, introducing complete dephasing between $|H
\rangle$ and $|V \rangle$, BS -- beam splitter, M -- mirror, PBS --
polarizing beamsplitter, SPD -- single photon detector, and FC --
fiber coupler. An ultraviolet doubled femtosecond pulsed laser
(about 100 mW, 390 nm, 76 MHz) is used to pump two BBO crystals for
type I down conversion to generate entangled photon pairs (780 nm,
$\frac{1}{\sqrt 2} (|HH\rangle -|VV\rangle)$), and the 2nd path is
split into two branches, one of them is controlled by optical
devices to prepare either a Bell-diagonal state with $b=0.8$ or a
Werner state with $q=0.5$. More precisely, we use an HWP to exchange
$|H\rangle$ and $|V\rangle$ on the branch to prepare the
Bell-diagonal state, while we use an HWP (transforming $|H\rangle$
and $|V\rangle$ to $|+\rangle$ and $|-\rangle$, respectively) and
QPs on the branch to prepare the Werner state. An attenuator is used
to tune the ration $q$ and $b$. The final state is detected by the
automatic tomography system, which is controlled by a LabVIEW
program and works precisely and efficiently.}
\end{figure}

\section*{III. Experimental Demonstration}

Now it is clear that estimation by $6^n$ measurements is better than
that by $4^n$. Next, we experimentally demonstrate that a complete
symmetrical set of measurement bases is the best choice for quantum
state estimation. Taking two qubits as an example, here we conduct
two all optical experiments on a Bell-diagonal state and a Werner
state to find the optimal number of measurements for quantum state
tomography. The experimental setup is described in FIG. 3, where the
polarization $|H \rangle$ and $|V \rangle$ are used to represent $|0
\rangle$ and $|1 \rangle$. Approximately $2.5\times 10^5$ photon
pairs are used in total, which are divided equally into $m=16$ to
$m=40$ measurement bases for tomography (see Table 1 for the
selection order of measurement bases). The results are shown in FIG.
4. They indicate that, among the tested sets of measurement bases,
the set containing 36 bases (an over-complete symmetrical set), is
the best; the others could not achieve such precision because they
are not distributed symmetrically. This illustrates once more the
importance of the choice of measurement bases.

\begin{figure}[tb]
\centering
\includegraphics[width=0.4\textwidth]{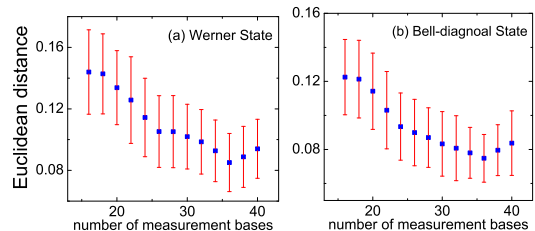}
\caption{\label{Fig:4}(color online) Error of different numbers of
measurement bases for $N=2.5 \times 10^5$. (a). Werner state with
$q=0.5$, (b). Bell-diagonal state with $b=0.8$. The $x$-axis shows
the number of measurement bases; the $y$-axis shows the Euclidean
distance. The state estimated from $10^8$ data is regarded as the
true one. We statistically calculate the mean value and standard
deviation of the Euclidean distance from 60 individual experiments
for each point, and we find that the mean value and standard
deviation have already converged to certain values. For both states,
the 36-base set is the best scheme.}
\end{figure}

In this experiment, the quantity of total resources is regarded as
the sum of used resources for all measurement bases. In fact, in
many experiments \cite{y5}, two orthogonal bases, such as $|
0\rangle$ and $| 1\rangle$, were measured simultaneously because the
PBS only separates them without degrading either one. In this way,
the total amount of resources used by over-complete QST can be saved
exponentially, and only $3^n$ measurements are required. Normal QST,
on the other hand, with its incomplete measurement bases, will waste
part of these resources (it also needs at least $3^n$ measurements).
One caution is necessary here, however, saving resources in this way
requires calibrating single photon detectors. Although a recent
research \cite{prl} introduces a novel set of mutually unbiased
bases for two qubits tomography, however, some of them are entangled
states and require more complex setting to detect. An over-complete
set of measurement bases in our experiment is easy to be realized in
experiments and can be extended to more than two qubits system.

\section*{IV. Conclusion}

In conclusion, we have made it clear that the precision of quantum
state estimation is strongly affected by the selection of
measurement bases. It has been proven that, when estimates are
performed with the same quantity of resources and the same
estimation method (MLE), over-complete QST makes a great difference
in improving the precision. Recently, several improved state
estimation methods have been proposed, such as Bayesian mean
estimation (BME) \cite{x1}, hedged maximum likelihood estimation
(HMLE) \cite{x2}, compressed sensing method \cite{x3}, and Marcus
Cramer et al.'s method \cite{x4} for finitely correlated states
(FCS) \cite{x5} or matrix product states (MPS) \cite{x6}. However,
these methods can be only used in some special situations, that is,
when the states are belong to some certain categories or prior
assumption of the states are available. They have not revealed the
intrinsic weakness of QST, the asymmetry distribution of the
measurement bases. Remarkably, over-complete QST can be combined
with these estimation methods to improve their efficiency
significantly. We can also try other symmetrical measurement bases,
such as the eight vertexes of the inscribed cube of the Bloch
sphere. This choice of bases is similar to the bases of
over-complete QST, and the result should also be very satisfactory.

Acknowledgment

This work was supported by the National Fundamental Research
Program, and the National Natural Science Foundation of China (Grant
Nos. 60921091, 10874162).

\begin{table*}[htbp]
\caption{CQST measurement bases selected for both experiments and
simulations. Here, $|L \rangle = \frac{1}{\sqrt 2} (|H \rangle + i|V
\rangle)$, $|R \rangle = \frac{1}{\sqrt 2} (|H \rangle - i|V
\rangle)$, $|+ \rangle = \frac{1}{\sqrt 2} (|H \rangle + |V
\rangle)$, and $|- \rangle = \frac{1}{\sqrt 2} (|H \rangle - |V
\rangle)$.} \vspace*{0.1in} \centering { \footnotesize{
\begin{tabular}{cccc}
\toprule%
$\nu$ & measurement base & $\nu$ & measurement base \\
\hline
1 & $|HH \rangle$ & 2 & $|HV \rangle$ \\
3 & $|VH \rangle$ & 4 & $|VV \rangle$ \\
5 & $|RH \rangle$ & 6 & $|RV \rangle$ \\
7 & $|+V \rangle$ & 8 & $|+H \rangle$ \\
9 & $|+R \rangle$ & 10 & $|++ \rangle$ \\
11 & $|R+ \rangle$ & 12 & $|H+ \rangle$ \\
13 & $|V+ \rangle$ & 14 & $|VL \rangle$ \\
15 & $|HL \rangle$ & 16 & $|RL \rangle$ \\
17 & $|LL \rangle$ & 18 & $|-L \rangle$ \\
19 & $|-+ \rangle$ & 20 & $|-- \rangle$ \\
21 & $|+- \rangle$ & 22 & $|L+ \rangle$ \\
23 & $|LR \rangle$ & 24 & $|-R \rangle$ \\
25 & $|HR \rangle$ & 26 & $|VR \rangle$ \\
27 & $|RR \rangle$ & 28 & $|+L \rangle$ \\
29 & $|-H \rangle$ & 30 & $|-V \rangle$ \\
31 & $|LH \rangle$ & 32 & $|LV \rangle$ \\
33 & $|H+ \rangle$ & 34 & $|V+ \rangle$ \\
35 & $|L+ \rangle$ & 36 & $|R+ \rangle$ \\
\end{tabular}
} } \label{Tab.1}
\end{table*}

\end{document}